\documentclass[14pt]{revtex4}

\usepackage{graphicx,epstopdf}

\begin{document}

\title{Quintom cosmological model and some possible solutions using Lie and Noether symmetries}
\author{Sourav Dutta$^1$\footnote {sduttaju@gmail.com}}
\author{Muthusamy Lakshmanan $^2$\footnote {lakshman.cnld@gmail.com}}
\author{Subenoy Chakraborty$^1$\footnote {schakraborty.math@gmail.com}}
\affiliation{$^1$Department of Mathematics, Jadavpur University, Kolkata-700032, West Bengal, India\\
$^2$ Centre for Nonlinear Dynamics, Bharathidasan University, Tiruchirapalli - 620 024, India}


\begin{abstract}
The present work deals with a quintom model of dark energy in  the framework of a spatially flat isotropic and homogeneous Friedmann--Lemaitre--Robertson--Walker (FLRW) universe. At first, Lie point symmetry is imposed to the system and the unknown coupled potential of the model is determined. Then Noether symmetry, which is also a point like symmetry of the Lagrangian, is imposed on the physical system and the potential takes a general form. It is shown that the Lie algebra of Noether symmetry is a sub-algebra of the corresponding Lie algebra of the Lie symmetry. Finally, a point transformation in the three dimensional augmented space is performed suitably so that one of the variables become cyclic and as a result there is considerable simplification to the physical system. Hence conserved quantities (i.e, constants of motion) are expressed in a compact form and cosmological solutions are evaluated and analyzed in the present context.\\\\


\end{abstract}
\maketitle

\section{introduction}

Modern cosmology at present is facing the challenging issue of dark energy which is supposed to be responsible for the present accelerating phase of the universe. The important observational evidences in support of this argument are based on the experimental study of (i) Supernovae Ia (SNIa) [1--3], (ii) cosmic microwave background (CMB) radiation along with large--scale structure surveys (LSS) [4, 5, 6],(iii) baryon acoustic oscillation (BAO) [7, 8], (iv) weak lensing [9] and (v) the age of the universe  calculated by introducing dark energy [10]. In fact, the above observational evidences are nicely accommodated in the frame work of Einstein's general theory of relativity by introducing an exotic matter [11--13] having large negative pressure --the dark energy (DE) for which the simplest explanation in agrement with the observations is the cosmological constant [14--19]. But due to its two severe drawbacks (namely cosmological constant problem and coincidence problem [20]) dynamical dark energy models [21--24] are widely used in the literature. However, most of these dynamical DE models describe the evolution of the universe not only in the quintessence era where the strong energy conditions are violated but the weak energy conditions are satisfied, i.e, $\rho+3p<0$ and $\rho+p>0$, but also in the phantom domain where all the energy conditions are violated, i.e, $\rho+p<0$. \\\\

 In the result of Planck' 2013 data [25] together with Supernovae data, the phantom region was favored at 2$\sigma$ level due to the small tension between planck and Supernovae data. However in the recent results of Planck collaboration [26], namely Planck' 2015 data set, this discrepancy is no longer present and the central value of the dark energy equation of state parameter is very close to -1. But this does not mean that the phantom region is ruled out because it is still compatible with the observational data within their experimental errors. Hence the equation of state parameter for dark energy $(w_d)$ may cross the cosmological constant boundary (i.e, $w_d$ may evolve from $w_d>-1$ to $w_d<-1$). On the other hand there exists a no--go theorem [27] which forbids the equation of state of a single scalar field to cross over the cosmological constant boundary. So a natural dark energy model in the perspective of recent Planck data consists of multi--dynamical scalar fields, particularly one which is a canonical scalar field and the other a phantom scalar field. Such phenomenological models are known as quintom models for which the dynamics of dark energy is described by parametrization of its equation of state [28--33]. Also it has been shown [34] that to violate the no--go theorem, i.e, to have a crossing at $w=-1$, it is necessary to have more than one degree of freedom-- a theoretical justification in favor of quintom paradigm.\\\\

In this connection, geometrical symmetries related to space--time, namely Lie point and Noether symmetries, have useful applications to physical problems. In particular, the conserved quantities (known as conserved charges) related to Noether symmetries can be chosen as a selection criterion to discriminate similar physical processes, for example several uses in dark energy models [35--36, 38--43]. It is worthy to mention that although we have listed several uses of Lie point symmetries in dark energy models, it would be relevant to highlight that it has been specifically applied to the study of phantom fields [41], which is a similar situation to the one under analysis in this work. On the other hand , from the mathematical point of view , the first integral or Noether integral in the context of Lie or Noether symmetry provides a tool to simplify a given system of differential equations or to determine the integrability of the system. Further one can check the self--consistency of phenomenological physical models using Noether symmetries. Moreover, it is possible to constrain the physical parameters involved in a system by imposing the symmetry [44] (for example potential of a scalar field, equation of state parameter for a fluid system, etc.). Also, we should like to mention that in the recent past there are lot of works [45--51] related to the above symmetries in physical systems moving in Riemannian spaces.\\\\

The present work is to study the quintom cosmology in the perspective of Lie and Noether symmetries and it is motivated from the following perspectives: At first, the coupled potential function for the two scalar fields is determined from the geometrical principles, rather than a phenomenological choice. Then cosmological solutions for the present quintom model are evaluated using the symmetries and the solutions are analyzed from physical perspectives.\\

 The paper is organized as follows: Section 2 deals with the basic equations related to quintom cosmological model. Application of Lie point symmetry to this model and consequences are presented in Section 3. Then Noether symmetry has been imposed to the physical system in Section 4 and different cosmological solutions are obtained and examined in the present perspective. The paper ends with a short discussion in Section 5.\\

 \section{Basic Equations in Quintom cosmology}

 This section is devoted to study the cosmological evolution in late phase using quintom model. In this model we have a canonical scalar field $\phi$ and a ghost field $\sigma$. Note that in comparison to a canonical scalar field, the ghost field has kinetic energy with negative sign and also normally have states of negative norm. The potential function $V(\phi, \sigma)$ of the field is an unknown function of the model. The action integral of the model takes the form (choosing the units : $8\pi G=1=c$)

 \begin{equation}
 A=\frac{1}{2}\int d^4 x \sqrt{-g}\big[R-g^{\mu \nu} \partial_{\mu} \phi \partial_{\gamma} \phi+g^{\mu \nu} \partial_{\mu} \sigma \partial_{\gamma} \sigma+V(\phi, \sigma)\big].
 \end{equation}

 For spatially flat (spatial curvature of the 3--space is zero due to observational data) Friedmann--Lemaitre--Robertson--Walker (FLRW) space--time, the line element is given by

 \begin{equation}
 ds^2=-dt^2+a^2(t)\big[dr^2+r^2 d\Omega_2^2\big],
 \end{equation}

 and the above point--like Lagrangian takes the form ( $\mathcal{L}$ is usually used for Lagrangian density but here $L$ is used for point like Lagrangian)

 \begin{equation}
 L\big(a, \dot{a}, \phi, \dot{\phi}, \sigma, \dot{\sigma}\big)=-3a \dot{a}^2+a^3\big(\frac{1}{2} \dot{\phi}^2-\frac{1}{2} \dot{\sigma}^2-V(\phi, \sigma)\big).
 \end{equation}

 Here $a(t)$ is known as the scale factor and $d\Omega_2^2=d\theta^2+sin^2 \theta d\phi^2$ is the metric on unit 2--sphere and the scalar fields are assumed to be homogeneous everywhere i.e, depending only on time. The Einstein field equations then take the explicit form

 \begin{equation}
 3\frac{\dot{a}^2}{a^2}=\rho=\frac{1}{2} \dot{\phi}^2-\frac{1}{2} \dot{\sigma}^2+V(\phi, \sigma),
 \end{equation}

 and
 \begin{equation}
 2\frac{\ddot{a}}{a}+\frac{\dot{a}^2}{a^2}=-p=-\frac{1}{2} \dot{\phi}^2+\frac{1}{2} \dot{\sigma}^2+V(\phi, \sigma),
 \end{equation}

 while the scalar fields have the evolution equations:

 \begin{equation}
 \ddot{\phi}+3H\dot{\phi}+\frac{\partial V}{\partial \phi}=0,
 \end{equation}

 and
 \begin{equation}
 \ddot{\sigma}+3H\dot{\sigma}-\frac{\partial V}{\partial \sigma}=0,
 \end{equation}

 where $H=\frac{\dot{a}}{a}$ is the usual Hubble parameter. It should be noted that out of the four evolution equations  (4)--(7) only three are independent and the other (usually (4)) is termed as the constraint equation.\\

 Further, the effective equation of state parameter is

 \begin{equation}
 w_{eff}=\frac{\rho}{p}=\frac{\dot{\phi}^2-\dot{\sigma}^2-V(\phi, \sigma)}{\dot{\phi}^2-\dot{\sigma}^2+V(\phi, \sigma)},
 \end{equation}

 and $q=-(1+\frac{\dot{H}}{H^2})$ is the deceleration parameter of the model. So the universe expands in an accelerated manner for $q<0$ while $q>0$ stands for decelerated expansion of the universe.\\

 For a feasible quintom model, it is necessary to have $w_{eff}(\neq 0)$ close to the cosmological constant boundary $w=-1$, so that the dynamical evolution of both scalar fields gives a quintom scenario with a smooth journey across the boundary $w=-1$. Now differentiating equation (8) and using the evolution equations (6) and (7) of the scalar fields we obtain

 \begin{equation}
 \frac{dw_{eff}}{dt}=\frac{-24 VH(\dot{\phi}^2-\dot{\sigma}^2)-4(\dot{\phi}^2-\dot{\sigma}^2+2V)(\dot{\phi}V_{\phi}+\dot{\sigma}V_{\sigma})}{(\dot{\phi}^2-\dot{\sigma}^2+2V)^2}.
\end{equation}

Usually, the above equation can be used as a consistency check when it is not possible to have any closed form of solution.  The above coupled system of non--linear ordinary differential equations (5)--(7) are very difficult to solve with conventional techniques. Also one has to choose $V(\phi, \sigma)$ purely phenomenologically. However, by imposing point symmetry (Lie/Noether) on them one can not only determine an expression for the potential function but can also be able to evaluate solutions to the system of differential equations. Further, Lie point symmetries can be considered as a tool for examining self consistency of phenomenological physical models. If the constituent equations of the symmetry conditions are not consistent then the phenomenological model cannot be considered as  self consistent. In the following sections we shall discuss possible solutions using the Lie and Noether symmetries to the physical system.

\section{Lie point symmetry and possible solutions:}

A system of second order ordinary differential equations of the form

\begin{equation}
\ddot{x}^i=w^i(t, x^j, \dot{x}^j),
\end{equation}

is said to have the vector field $X=\xi \frac{\partial}{\partial t}+\eta^i \frac{\partial}{\partial x^i}$ in the augmented space $(t, x^i)$  (where the augmented space is the space of the both independent and dependent variables)as the generator of a Lie point symmetry [47] provided

\begin{equation}
X^{[2]}\big(\ddot{x}^i-w^i(t, x^j, \ddot{x}^j)\big)=0,
\end{equation}

here $X^{[2]}=\xi \partial_t+\eta^i \partial_{x^i}+\big(\dot{\eta}^i-\dot{x}^i \dot{\xi}\big)\partial \dot{x}^i+\left(\ddot{\eta}^i-\dot{x}^i\ddot{\xi}-2\ddot{x}^2\dot{\xi}\right)\partial \ddot{x}^i$ is termed as second prolongation of generator $X$. One can then define the jet space as an extension of the augmented space where the co-ordinates consist of not only  the independent and dependent variables but also the partial derivatives of the dependent variables. As in the present problem we shall deal with second order differential equations so that the jet space has upto second order derivatives as its co-ordinates and the prolongated vector field $X^{[2]}$ is nothing but the extension of the infinitesimal generator $X$ over the jet space. Equivalently, the above symmetry condition can be written as

\begin{equation}
\big[X^{[1]}, A\big]=\lambda(x^i) A,
\end{equation}

here $X^{[1]}$ the first prolongation of the vector field $X$ has the expression

\begin{equation}
X^{[1]}=\xi \partial_t+\eta^i \partial_{x^i}+\big(\dot{\eta}^i-\dot{x}^i \dot{\xi}\big)\partial \dot{x}^i,
\end{equation}

and
\begin{equation}
A=\partial_t+\dot{x}^i \partial_{x^{i}}+w^i\big(t, x^j, \dot{x}^j\big)\partial_{\dot{x}^i},
\end{equation}

is termed as the Hamiltonian vector field.\\

For the present quintom cosmological model, the augmented space is a four dimensional space: $(t, a, \phi, \sigma)$ and the evolution equations (5)--(7) can be written suitably as

\begin{equation}
\ddot{a}=\frac{\dot{a}^2}{a}-\frac{a}{2}\big(\dot{\phi}^2-\dot{\sigma}^2\big)\equiv w_1\big(t, a, \dot{a}, \phi, \dot{\phi}, \sigma, \dot{\sigma}\big),
\end{equation}

\begin{equation}
\ddot{\phi}=-3\frac{\dot{a}}{a} \dot{\phi}-\frac{\partial V}{\partial \phi}\equiv w_2\big(t, a, \dot{a}, \phi, \dot{\phi}, \sigma, \dot{\sigma}\big),
\end{equation}

\begin{equation}
\ddot{\sigma}=-3\frac{\dot{a}}{a} \dot{\sigma}+\frac{\partial V}{\partial \phi}\equiv w_3\big(t, a, \dot{a}, \phi, \dot{\phi}, \sigma. \dot{\sigma}\big).
\end{equation}

Hence the infinitesimal generator of the point transformation corresponding to equations (15)--(17) can be written as

\begin{equation}
X=\xi \frac{\partial}{\partial t}+\eta_1 \frac{\partial}{\partial a}+\eta_2 \frac{\partial}{\partial \phi}+\eta_3 \frac{\partial}{\partial \sigma}+\eta'_1 \frac{\partial}{\partial \dot{a}}+\eta'_2 \frac{\partial}{\partial \dot{\phi}}+\eta'_3 \frac{\partial}{\partial \dot{\sigma}},
\end{equation}

where $\xi,~\eta_i(i=1, 2, 3)$ are functions in the augmented space, i.e, $\xi=\xi(t, a, \phi, \sigma),~\eta_i=\eta_i(t, a, \phi, \sigma).$ Also $\eta'_i$ are defined as

\begin{equation}
\eta'_i=\frac{d \eta_i}{dt}-a_i\frac{d \xi}{dt},~~~~i=1, 2, 3
\end{equation}

with $a_i\equiv(a, \phi, \sigma)$ for $i=(1, 2, 3)$ and the total derivative operator $\frac{d}{dt}$ has the expression

\begin{equation}
\frac{d}{dt}=\frac{\partial}{\partial t}+\dot{a}\frac{\partial}{\partial a}+\dot{\phi}\frac{\partial}{\partial \phi}+\dot{\sigma}\frac{\partial}{\partial \sigma}.
\end{equation}

Now the Lie point symmetry conditions corresponding to the evolution equations (15)--(17) can be written as [47]

\begin{equation}
Xw_1=\eta''_1,~~Xw_2=\eta''_2,~~Xw_3=\eta''_3,
\end{equation}

where
\begin{equation}
\eta_i^{\prime\prime}=\frac{d\eta_i^{\prime}}{dt}-\ddot{a_i}\frac{d\xi}{dt}=\frac{d\eta_i^{\prime}}{dt}-w_i\frac{d\xi}{dt},~~i=1, 2, 3.
\end{equation}

So corresponding to the symmetry conditions (21), we have a set of overdetermined equations whose solutions give the components of the symmetry as

\begin{equation}
\xi=\xi_0,~\eta_1=\eta_{01} a, \eta_2=\eta_{02},~\eta_3=\eta_{03},
\end{equation}

 with~$V(\phi, \sigma)=V_0(\phi-\sigma)$, where $\xi_0, \eta_{01}, \eta_{02}, \eta_{03}$ and $V_0$ are arbitrary constants. In particular if we choose $\xi_0=0=\eta_{01}$ then the corresponding Lie symmetry will be similar to the Noether symmetry that we shall describe in the next section.\\

So the symmetry vector can be written as

\begin{equation}
X=\xi_0 \partial_t+\eta_{01} a\frac{\partial}{\partial a}+\eta_{02} \frac{\partial}{\partial \phi}+\eta_{03} \frac{\partial}{\partial \sigma},
\end{equation}

and as a result the characteristic equation takes the form

\begin{equation}
\frac{dt}{\xi_0}=\frac{da}{\eta_{01} a}=\frac{d\phi}{\eta_{02}}=\frac{d \sigma}{\eta_{03}}.
\end{equation}

Hence we have the three parameter family of functions

\begin{equation}
\phi=\frac{\eta_{02}}{\xi_0}t+\phi_0,~\sigma=\frac{\eta_{03}}{\xi_0}t+\sigma_0,~a=a_0 e^{\frac{\eta_{01}}{\xi_0}t},
\end{equation}
where $a_0, \phi_0$ and $\sigma_0$ are the constants of integration.\\

Using this parametric family of functions to the field equation (4)--(7) the relations among the constants can be obtained as follows:

\begin{equation}
\eta_{02}=-\eta_{03}, \mbox{and}~3\eta_{01} \eta_{03}=V_0 \xi_0^2.
\end{equation}

\section{Noether symmetry approach in quintom cosmology}

According to Noether's first theorem, any physical system is associated to some conserved quantities if the Lagrangian of the system is invariant with respect to the Lie derivative [52] along an appropriate vector field ($\mathcal{L}_{\overrightarrow{V}}f=\overrightarrow{V}(f)$). Also by imposing these symmetry constraints, the evolution equations of the physical system are either solvable or simplified to a great extent.\\

The Euler--Lagrange equations
\begin{equation}
\partial_j\big(\frac{\partial L}{\partial \partial_j q^{\alpha}}\big)=\frac{\partial L}{\partial q^{\alpha}},~\alpha=1, 2,......N
\end{equation}

corresponding to a point--like canonical Lagrangian

\begin{equation}
L=L\big[q^{\alpha}(x^i), \partial_j q^{\alpha}(x^i)\big],
\end{equation}

with generalized co--ordinate $q^{\alpha}(x^i)$, contracting with some unknown function $\lambda^\alpha(q^\beta),$ gives

\begin{equation}
\lambda^\alpha\left[\partial_j\left(\frac{\partial L}{\partial \partial_j q^\alpha}\right)-\frac{\partial L}{\partial q^\alpha}\right]=0,
\end{equation}

i.e,
\begin{equation}
\mathcal{L}_{\overrightarrow{X}}L=\overrightarrow{X}(L)=\lambda^\alpha\frac{\partial L}{\partial q^\alpha}+(\partial_j \lambda^{\alpha})\frac{\partial L}{\partial \partial_j q^\alpha}=\partial_j\left(\lambda^\alpha\frac{\partial L}{\partial \partial_j q^\alpha}\right).
\end{equation}

Here\begin{equation}
\overrightarrow{X}=\lambda^\alpha\frac{\partial}{\partial q^\alpha}+\left(\partial_j\lambda^\alpha\right)\frac{\partial}{\partial \partial_j q^\alpha},
\end{equation}

is the vector field with respect to which Lie derivative  [52] is considered. Now, if it so happens that the vector field $\overrightarrow{X}$ is the infinitesimal generator of Noether symmetry, then by Noether's theorem $\mathcal{L}_{\overrightarrow{X}}L=0$ and we have from equation (31), a conserved current corresponding to this symmetry as  [53]

\begin{equation}
Q^i=\lambda^\alpha\frac{\partial L}{\partial \partial_i q^\alpha},
\end{equation}

i.e, $\partial_iQ^i=0.$\\

The energy function associated with a given Lagrangian is given by

\begin{equation}
E=\frac{\partial L}{\partial \dot{q}^\alpha}\dot{q}^\alpha-L.
\end{equation}

 The energy function, which is better known as Hamiltonian, is a constant of motion only when there is no explicit time dependence in the Lagrangian [37]. Further, if due to symmetry the conserved quantity has some physical meaning [40] then the Noether symmetry approach can be used to select reliable models. In the following we shall show how by using the Noether symmetries the evolution equations can be reduced and the physical problem can be solved exactly.\\

 For the present quintom model the generalized co-ordinate $(a, \phi, \sigma)$ are functions of time only so the Lagrangian $L$ in equation (29) is a function of time alone. Thus the Noether symmetry is generated by the infinitesimal generator

\begin{equation}
\overrightarrow{X}=\alpha\frac{\partial}{\partial a}+\beta\frac{\partial}{\partial \phi}+\gamma\frac{\partial}{\partial \sigma}+\dot{\alpha}\frac{\partial}{\partial \dot{a}}+\dot{\beta}\frac{\partial}{\partial \dot{\phi}}+\dot{\gamma}\frac{\partial}{\partial \dot{\sigma}},
\end{equation}

here the unknown coefficients $\alpha, \beta$ and $\gamma$ are functions of the dynamical variables $a, \phi$ and $\sigma$, i.e,

$$\alpha=\alpha(a, \phi, \sigma ),~~\beta=\beta(a, \phi, \sigma)~\mbox{and}~\gamma=\gamma(a, \phi, \sigma)$$
and we have
$$\dot{\alpha}=\frac{\partial \alpha}{\partial a}\dot{a}+\frac{\partial \alpha}{\partial \phi}\dot{\phi}+\frac{\partial \alpha}{\partial \sigma}\dot{\sigma},$$ and similarly for $\dot{\beta}$ and $\dot{\gamma}.$\\

Now by imposing Noether symmetry to the Lagrangian i.e, $\mathcal{L}_{\overrightarrow{X}}L=0,$ the coefficients of the infinitesimal generator will satisfy an over determined system of partial differential equations:

\begin{eqnarray}
2a \alpha_a+\alpha&=&0,~2a \beta_{\phi}+3\alpha=0,~2a \gamma_{\sigma}+3\alpha=0,
\nonumber
\\
a^2\beta_a-6 \alpha_{\phi}&=&0,~a^2\gamma_a+6\alpha_{\sigma}=0,~\beta_{\sigma}=\gamma_{\phi},~\beta_{\phi}=\gamma_{\sigma},
\end{eqnarray}

with an additional equation for the potential
 \begin{equation}
 a\big(\beta V_{\phi}+\gamma V_{\sigma}\big)+3\alpha V=0.
 \end{equation}

 In the above we have notationally denoted the partial derivative by a suffix (i.e, $\alpha_a\equiv \frac{\partial \alpha}{\partial a}$).\\

 The standard method of separation of variables for solving the above set of partial differential equations is employed and hence we write

 \begin{eqnarray}
\alpha&\equiv&\alpha (a, \phi, \sigma)= \alpha_1 (a) \alpha_2 (\phi) \alpha_3 (\sigma),
\nonumber
\\
\beta&\equiv&\beta (a, \phi, \sigma)= \beta_1 (a) \beta_2 (\phi) \beta_3 (\sigma),
\nonumber
\\
\gamma&\equiv&\gamma (a, \phi, \sigma)= \gamma_1 (a) \gamma_2 (\phi) \gamma_3 (\sigma).
\end{eqnarray}

The explicit solution gives

\begin{equation}
\alpha=0,~\beta=k\sigma+d_1,~\gamma=k\phi+d_2,
\end{equation}

where $k, d_1, d_2$ are arbitrary integration constants.\\
Also the unknown potential function has the expression

\begin{equation}
V(\phi, \sigma)=v_1 \big[\frac{k}{2}\big(\phi^2-\sigma^2 \big)+d_2 \phi-d_1 \sigma \big],
\end{equation}

with $v_1,$ a constant. Thus the potential function is a polynomial function in the field variables $(\phi, \sigma)$. Note that if $k=0, d_1=d_2$ then the potential function will be invariant under the linear shift: $\phi \rightarrow \phi+l, \sigma \rightarrow \sigma+l$ ($l$ is arbitrary). The potential will have the reflection symmetry (i.e, $\phi \rightarrow -\phi, \sigma \rightarrow -\sigma$) if $d_1=0=d_2$. Further, if for $d_1=0=d_2$ we choose the potential to be \\
  $$V(\phi, \sigma)=v_1 \frac{k}{2}\big(|\phi|^2-|\sigma|^2\big)$$
  then $V$ is rotational invariant, i.e, under the transformation $\phi \rightarrow \phi e^{i\theta_1}, \sigma \rightarrow \sigma e^{i\theta_2}, (\theta_1, \theta_2$ are arbitrary). As we are considering only real scalar field so in the present context potential does not have rotational invariant.\\

   It should be noted that this Noether symmetry is similar to the Lie symmetry that we have obtained in the previous section with the particular choice of $\xi_0=0=\eta_{01}$. Hence we can say that the corresponding Lie algebra of Noether symmetry is a subalgebra of the Lie algebra of the Lie symmetry obtained in the previous section.\\

Thus, the infinitesimal generator corresponding to Noether symmetry is completely determined (except for arbitrary constants) and the expression for potential function has been derived in the course of their determination.\\

 It is important to note that in general, for a field theory in curved space, there is no well-defined notion of energy. In fact, the conserved quantity derived from Noether's theorem is the energy--momentum tensor. However for the particular case in which there is a time like Killing vector, a conserved energy can be defined. This is not the case of FLRW metrics. But for the present model as the Lagrangian does not depend explicitly on time so in analogy with the point--like Lagrangian one can define an energy which will be conserved in nature (for general Lagrangian, the Hamiltonian is not necessarily conserved). Hence in the context of present Noether symmetry the two constants of motion, namely conserved charge (defined in equation(33)), and conserved energy (in equation(34)) have the explicit expressions (using equation (39))

\begin{equation}
Q=a^3 k\big(\sigma \dot{\phi}-\dot{\sigma} \phi \big)+a^3 \big( d_1 \dot{\phi}-d_2 \dot{\sigma}\big),
\end{equation}

and

\begin{equation}
E=-3a \dot{a}^2+\frac{1}{2}a^3 \dot{\phi}^2-\frac{1}{2} a^3\dot{\sigma}^2+a^3 V(\phi, \sigma).
\end{equation}

 Normally associated with Noether symmetry we have a conserved current as defined in equation(33). Its time component when integrated over the space volume gives a conserved charge. However, for the present case as noted above the Lagrangian depends only on the time co-ordinate  (i.e, generalized co-ordiantes depend only on time) and so $Q$ in equation  (41) represents the Noether charge.\\

We now introduce the Cartan one form [38]

\begin{equation}
\theta_L=\frac{\partial L}{\partial \dot{a}}da+\frac{\partial L}{\partial \dot{\phi}}d\phi+\frac{\partial L}{\partial \dot{\sigma}}d\sigma,
\end{equation}

so that the above conserved charge can be written as the inner product (i.e, contraction) between the vector field $\overrightarrow{X}$ and the differential form $\theta_L$, i.e,

\begin{figure}
\includegraphics[width=0.5\textwidth]{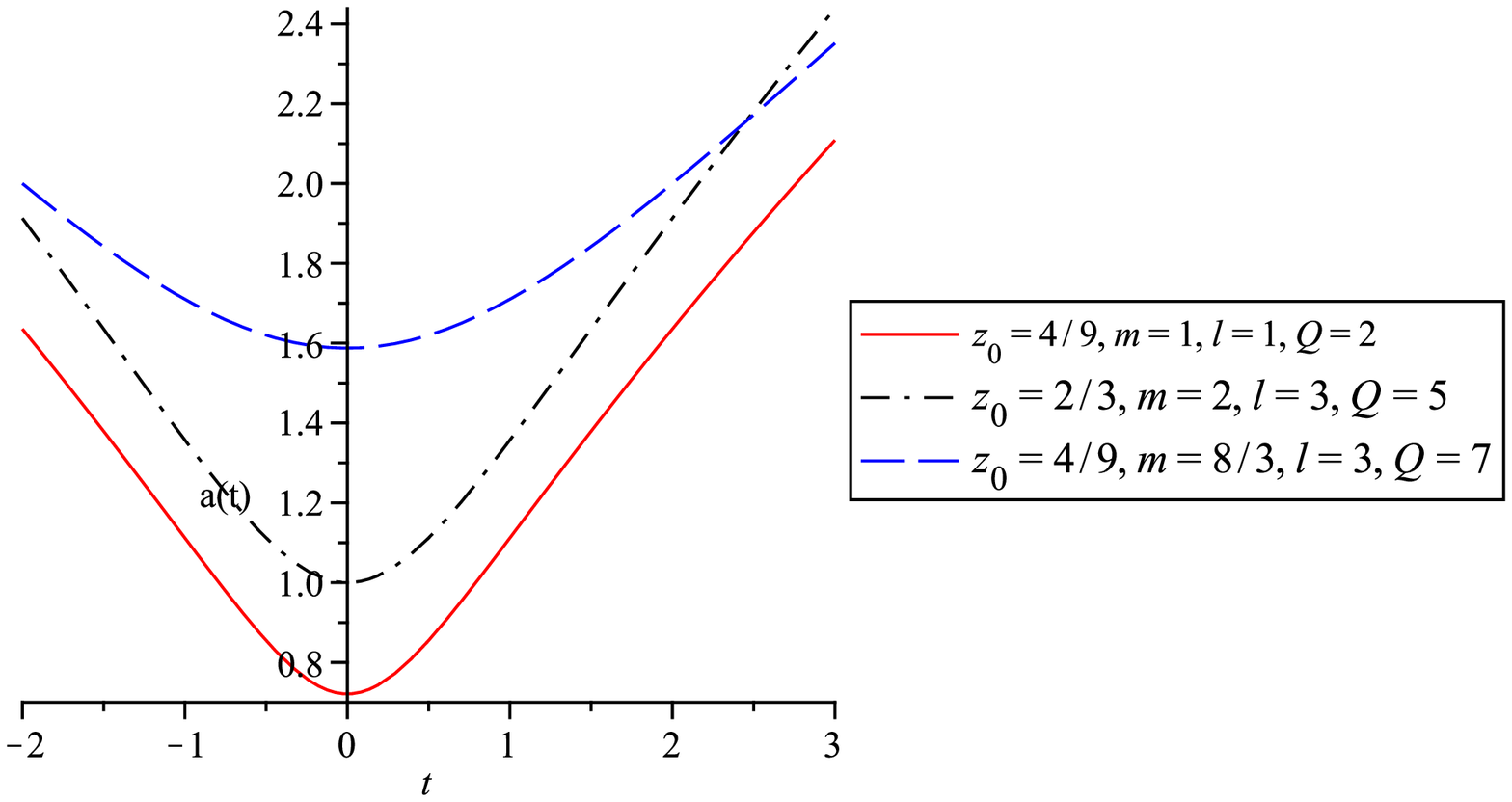}\\
Figure 1: The scale factor $a(t)$ against $t$ for $l< Q$.
\end{figure}
\begin{figure}
\includegraphics[width=0.5\textwidth]{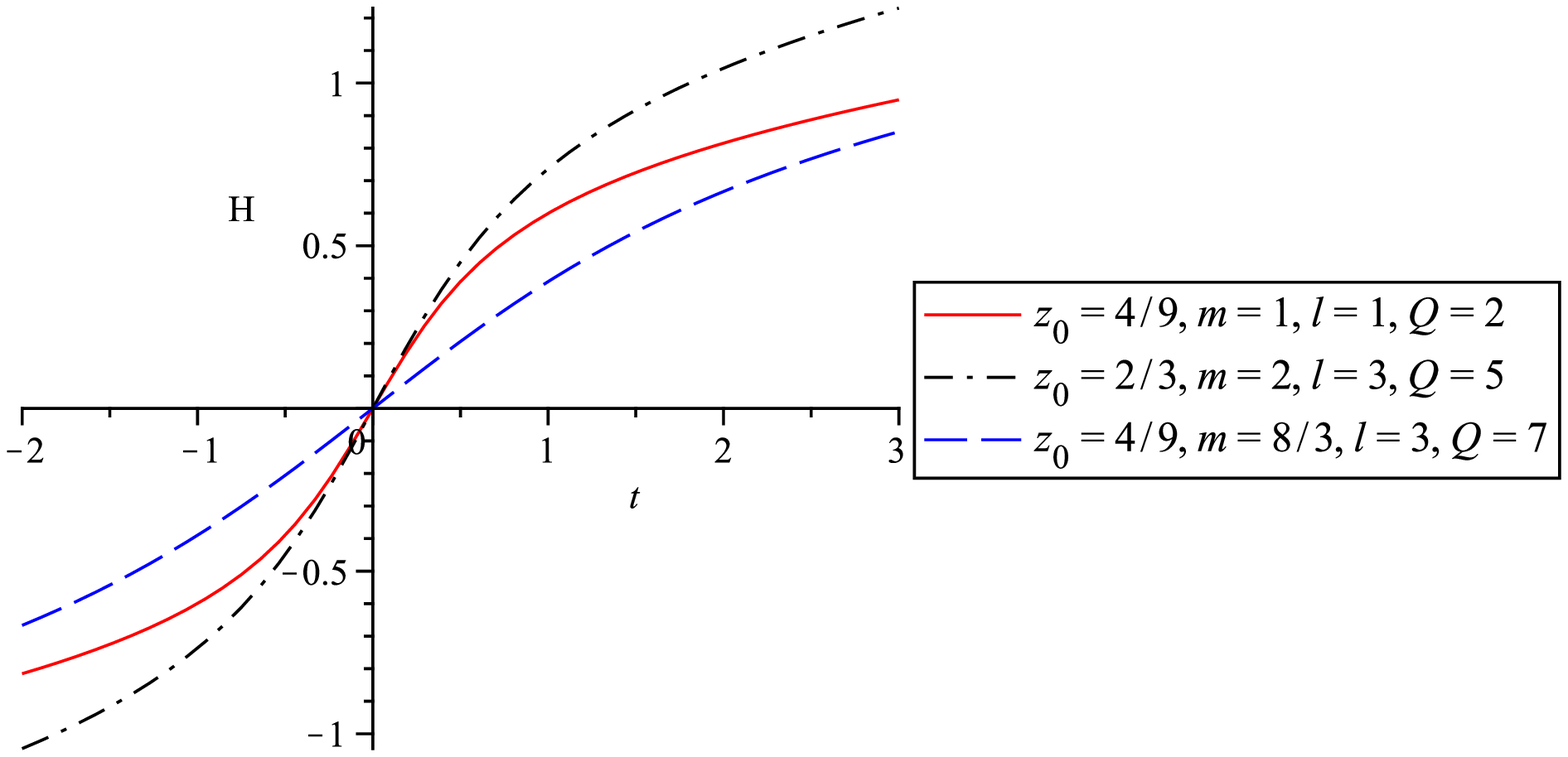}\\
Figure 2: Variation of Hubble parameter against $t$ for $l < Q$.
\end{figure}
\begin{figure}
\includegraphics[width=0.5\textwidth]{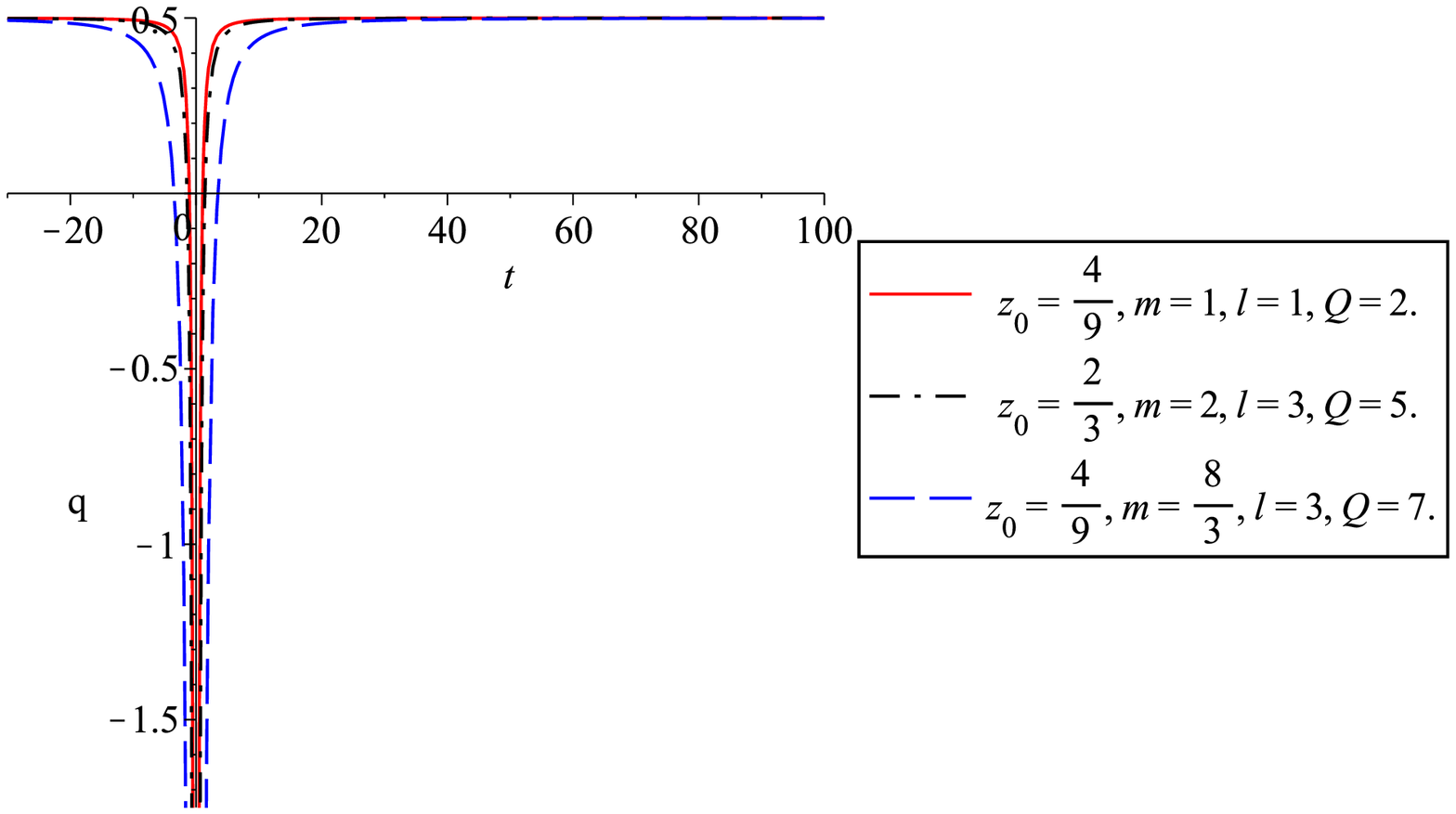}\\
Figure 3: Graphical representation of  $q$ against $t$ for $l < Q$.
\end{figure}

\begin{equation}
Q=i_{\overrightarrow{X}} \theta_L.
\end{equation}

where $i_{\overrightarrow{X}}$ stands for the inner product with the vector field $X$. This representation is useful to identify cyclic variables in the Lagrangian of the system. Note that a generalized co-ordinate which does not appear explicitly in the Lagrangian even though its time derivative is present in it is called a cyclic variable.\\

In order to simplify the evolution equations we make a transformation of variables in the augmented space i.e, $(a, \phi, \sigma)\rightarrow (u, v, w)$ so that one of the transformed variables (say $u$) becomes cyclic and hence the transformed Lagrangian becomes much simpler in form and as a result the evolution equations become solvable. The infinitesimal generator (i.e, the vector field $\overrightarrow{X}$), due to this point transformation : $(a, \phi, \sigma)\rightarrow (u, v, w),$ is transformed into

\begin{figure}
\includegraphics[width=0.5\textwidth]{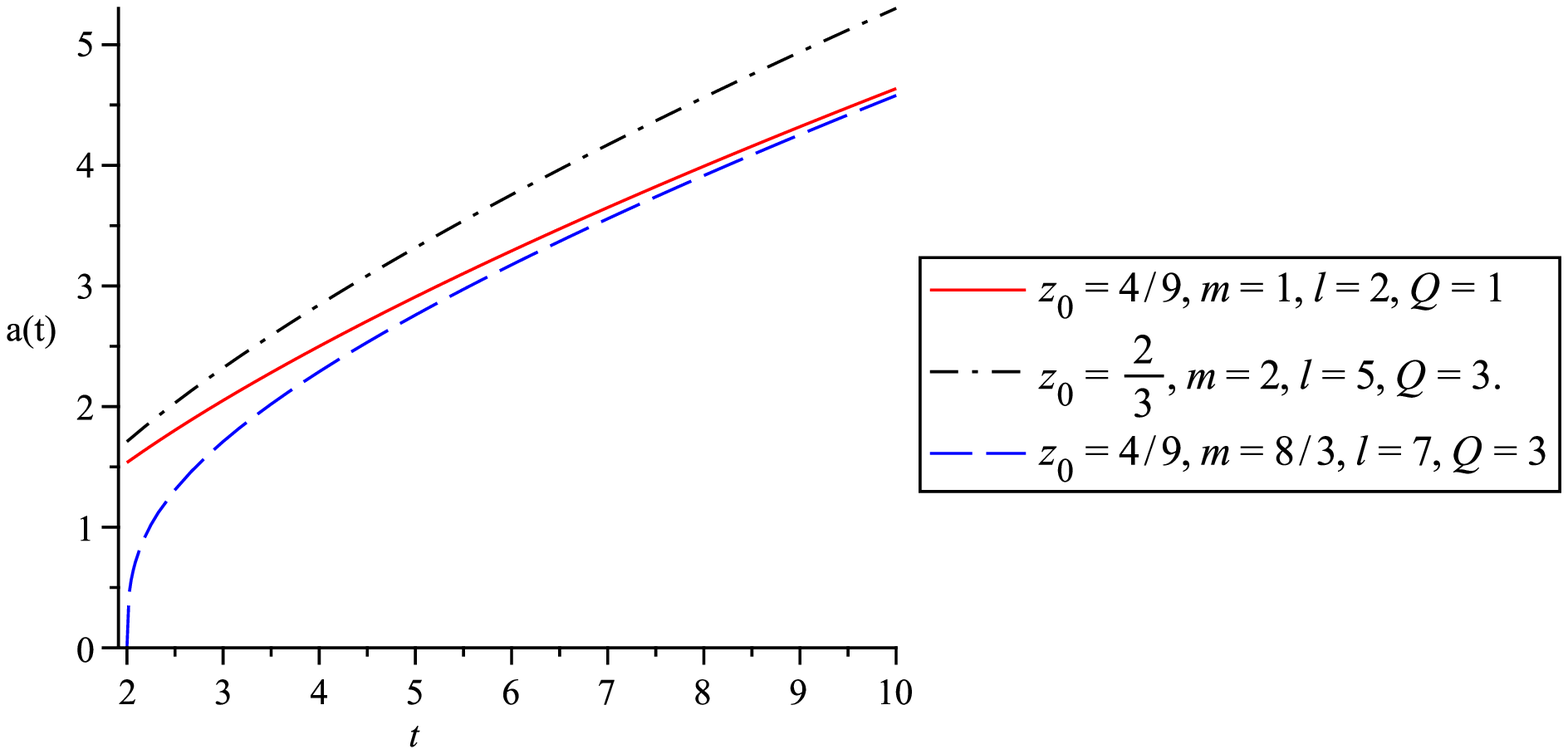}\\
Figure 4: Scale factor $a(t)$ against $t$ for $l>Q$.
\end{figure}
\begin{figure}
\includegraphics[width=0.5\textwidth]{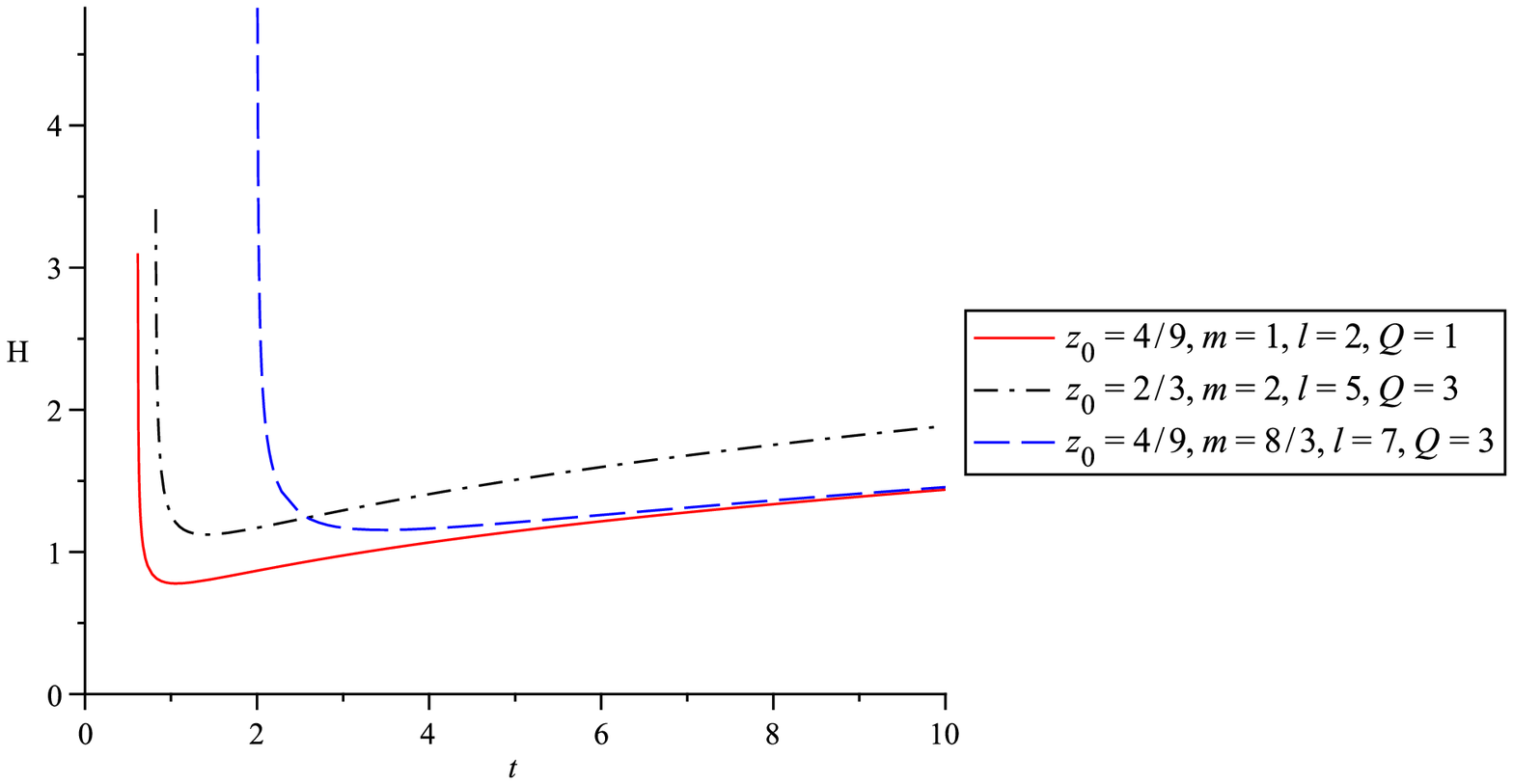}\\
Figure 5: Variation of Hubble parameter against $t$ for $l > Q$.
\end{figure}
\begin{figure}
\includegraphics[width=0.5\textwidth]{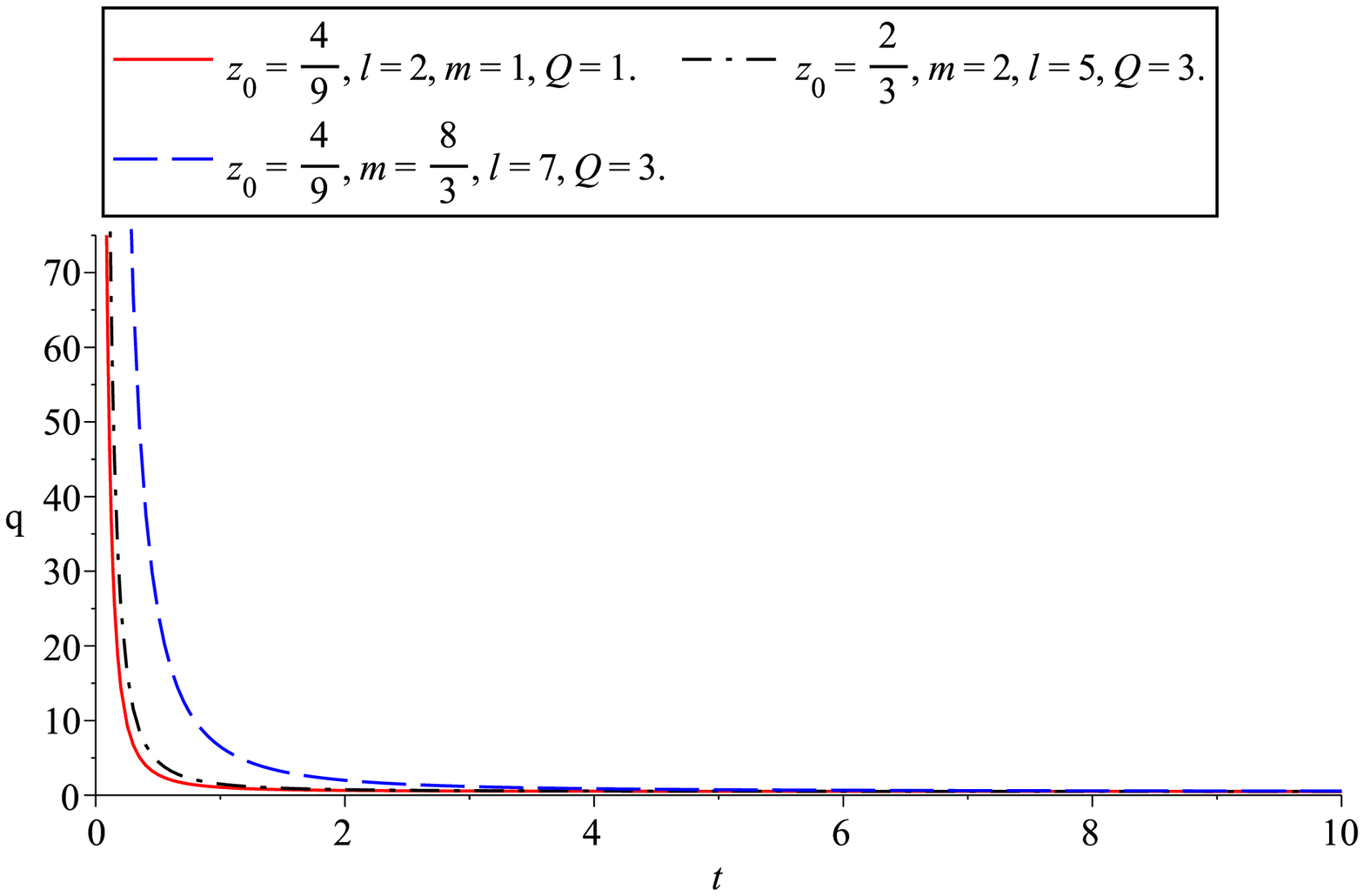}\\
Figure 6:  Graphical representation of  $q$ against $t$ for $l > Q$.
\end{figure}

\begin{equation}
\overrightarrow{X}_T=(i_{\overrightarrow{X}} du)\frac{\partial}{\partial u}+(i_{\overrightarrow{X}}dv)\frac{\partial}{\partial v}+(i_{\overrightarrow{X}} dw)\frac{\partial}{\partial w}+\left(\frac{d}{dt}(i_{\overrightarrow{X}}du)\right)\frac{d}{d \dot{u}}+\left(\frac{d}{dt}(i_{\overrightarrow{X}}dv)\right)\frac{d}{d \dot{v}}+\left(\frac{d}{dt}(i_{\overrightarrow{X}}dw)\right)\frac{d}{d \dot{w}}  .
\end{equation}

So one may consider $\overrightarrow{X}_T$ as the lift of a vector field defined on the augmented space. Further, we restrict the above point transformation (without any loss of generality) such that

\begin{equation}
i_{\overrightarrow{X}} du=1~~\mbox{and}~~ i_{\overrightarrow{X}}dv=0=i_{\overrightarrow{X}}dw,
\end{equation}

 and then
$${\overrightarrow{X_T}}=\frac{\partial}{\partial u},~\frac{\partial L_T}{\partial u}=0.$$

 It is evident that $u$ is the cyclic co-ordinate and so the dynamics can be reduced [54]. Further, the above restriction (i.e, equation (46)) for the point transformation is essentially choosing the transformed infinitesimal generator $\overrightarrow{X}_T$ along one co-ordinate line (say along $u$ co-ordinate). As a result, $\overrightarrow{X}_T$ is simplified to a great extent as $\frac{\partial}{\partial u}$ and hence the Lagrangian {\bf $L_T$} in terms of new variables becomes invariant under the translation with respect to $u$ (i.e, $u$ is a cyclic co--ordinate [41].)\\

It is easy to solve the first order linear partial differential equations in (46) to obtain the interrelation between the old and new variables in the augmented space as the following:

{\bf I: $k\neq 0$}

\begin{eqnarray}
e^{ku}&=& k\big(\phi+\sigma \big)d_1+d_2,
\nonumber
\\
v&=& a,
\nonumber
\\
w&=& \frac{k}{2}\big(\phi^2-\sigma^2 \big)+d_1 \phi-d_2 \sigma.
\end{eqnarray}

{\bf II: $k= 0$}

\begin{eqnarray}
u&=& \frac{\phi+\sigma}{d_1+d_2},
\nonumber
\\
v&=& a,
\nonumber
\\
w&=& d_1 \phi-d_2 \sigma.
\end{eqnarray}

So the Lagrangian in the transformed variables takes the form:

\begin{equation}
L_T=\left\{
\begin{array}{lll}
-3v\dot{v}^2+v^3\big(\dot{w} \dot{u}-w k \dot{u}^2-v_1 w\big), & \mbox{ for } k\neq 0, d_1=d_2=0\\
 -3v\dot{v}^2+v^3\big(\dot{w} \dot{u}-v_1 w\big), & \mbox{ for } k=0, d_1=d_2=d\\
-3v\dot{v}^2+v^3\big(\dot{w} \dot{u}-v_1 w\big) & \mbox{ for } k= 0, d_2=-d_1=d\\
\end{array}
\right\},
\end{equation}

Also the conserved quantities in new variables takes the form\\

{\bf (a):~\underline {for $k\neq 0,~d_1=d_2=0$}}
\begin{eqnarray}
Q&=&v^3\big(\dot{w}-2kw\dot{u}\big)\nonumber,\\
E&=&-3v\dot{v}^2+v^3\big(\dot{w} \dot{u}-kw\dot{u}^2\big)+v_1 v^3w,
\end{eqnarray}

{\bf (b):~\underline {for $k= 0,~d_1=d_2=d$}}
\begin{eqnarray}
Q&=&v^3\dot{w}\nonumber,\\
E&=& v^3\big(-3\frac{\dot{v}^2}{v^2}+\dot{w} \dot{u}+v_1w\big),
\end{eqnarray}

{\bf (c):~\underline {for $k= 0,~d_2=-d_1=d$}}
\begin{eqnarray}
Q&=&-v^3\dot{w}\nonumber,\\
E&=& v^3\big(-3\frac{\dot{v}^2}{v^2}-\dot{w} \dot{u}+v_1w\big).
\end{eqnarray}

Now solving the Euler--Lagrange equations for the new Lagrangian in equation (48) and using the above two first integrals we have the following solutions:\\

{\bf I:~\underline {for $k\neq 0,~d_1=d_2=0$}}

\begin{equation}
u=\left\{
\begin{array}{lll}
-\frac{4m}{9z_0 \mu} coth^{-1}(\frac{T}{\mu}),& \mbox{ for }l>Q,~\mu^2=\frac{2m}{27z_0^2}(l-Q)\\
 \frac{4m}{9z_0 \lambda} tan^{-1}(\frac{T}{\lambda}),& \mbox{ for }l<Q,~\lambda^2=\frac{2m}{27z_0^2}(Q-l)\\
-\frac{4m}{9z_0}\frac{1}{T},& \mbox{ for }l=Q,~T=t-t_0\\
\end{array}
\right\},
\end{equation}

\begin{equation}
v^3=\frac{9}{4}z_0T^2-\frac{m(l-Q)}{6z_0}
\end{equation}

\begin{equation}
w=\left\{
\begin{array}{lll}
-\frac{4(l-Q)}{9z_0u}coth^{-1}(\frac{T}{\mu}),& \mbox{ for } l>Q\\
\frac{4(Q-l)}{9z_0\lambda}cot^{-1}(\frac{T}{\lambda}),& \mbox{ for } l<Q\\
 w_0T,& \mbox{ for } l=Q\\
\end{array}
\right\},
\end{equation}

where $z_0, t_0$ and $w_0$ are constants of integration and $l$ and $m$ are the conserved momenta associated with the cyclic variables $u$ and $w$. Hence the solution for old variables are the following:

\begin{equation}
 a=v=\left[\frac{9}{4}z_0T^2-\frac{m(l-Q)}{6z_0}\right]^{\frac{1}{3}},
\end{equation}

\begin{equation}
\phi=\frac{1}{2}\left(U+\frac{2w}{kU}\right),~\sigma=\frac{1}{2}\left(U-\frac{2w}{kU}\right),~U=e^{ku}.
\end{equation}

{\bf II:~\underline {for $k =0,~d_1=d=d_2$}}

\begin{eqnarray}
u&=&p_0+\frac{(E+3u_0^2)}{Q^i}T-\frac{3v_1r_0u_0^2}{4Q^i}T^3+\frac{v_1}{2}T^2,\nonumber\\
v&=&\left(\frac{9u_0}{4}\right)^{\frac{1}{3}}T^{\frac{2}{3}},\nonumber\\
w&=&w_0-\frac{4Q}{9u_0^2}\frac{1}{T}.
\end{eqnarray}
Then the solutions for the old variables are

\begin{eqnarray}
a&=&\left(\frac{9u_0}{4}\right)^{\frac{1}{3}}T^{\frac{2}{3}},\nonumber\\
\phi&=&\frac{1}{2}\left(2 d u+\frac{w}{d}\right),\nonumber\\
\sigma&=&\frac{1}{2}\left(2 d u-\frac{w}{d}\right).
\end{eqnarray}

{\bf III:~\underline {for $k =0,~d_2=-d_1=d$}}

\begin{eqnarray}
u&=&p_0+\frac{(E+3u_0^2)}{Q}T-\frac{3v_1r_0u_0^2}{4Q}T^3-\frac{v_1}{2}T^2,\nonumber\\
v&=&\left(\frac{9u_0}{4}\right)^{\frac{1}{3}}T^{\frac{2}{3}},\nonumber\\
w&=&w_0+\frac{4Q}{9u_0^2}\frac{1}{T}.
\end{eqnarray}

Then explicit solutions for the old variables are

\begin{eqnarray}
a&=&\left(\frac{9u_0}{4}\right)^{\frac{1}{3}}T^{\frac{2}{3}},\nonumber\\
\phi&=&\frac{1}{2}\left(\frac{w}{d}-2ud\right),\nonumber\\
\sigma&=&\frac{1}{2}\left(2 d u+\frac{w}{d}\right).
\end{eqnarray}

\section{Summary}

The present paper studies a quintom model of dark energy in the background of flat FLRW model of the universe. The model is described and analyzed from the point of view of point symmetries namely Lie and Noether symmetries. Note that the Lie point symmetries of a differential equation form a Lie algebra. However, if the differential equations (i.e, evolution equations) are obtained from a variational principle i.e, they are the Euler--Lagrange equations of some Lagrangian function then Lie point symmetries which transform the action integral, keeping the Euler--Lagrange equations invariant, are termed as Noether point symmetries. The special feature of each Noether symmetry is that there should be a conservation law with an associated as Noether integral/Noether Invariant [55, 56]. Further, the set of all the Noether point symmetries of a differential equation form a subalgebra of the Lie point symmetries of that equation and is known as Noether algebra of the differential equation. \\

 The two scalar fields quintom model is considered as the simplest way to avoid the no--go criterion for the description of phantom era by scalar field models for DE. In most of the scalar field models describing DE, the potential is chosen phenomenologically as there is no physical basis for the choice of the potential. However one is able determine an analytic expression for the potential using symmetry conditions. In addition, some analytic solutions may be obtained from symmetry criteria. In the present work we are successful from both the perspectives, i.e, the potential for the quintom model is obtained in simple polynomial form of the scalar fields (equation (40)) using symmetry conditions and three sets of solutions (in equations (56), (58) and (60)) are obtain for the field equations with the help of conserved charge and conserved energy.\\

The solution (56) shows a bouncing model of the cosmic evolution for $l<Q$ as shown in figure 1 for different choices of the parameter involved. The contracting phase of the universe corresponds to $t<0$ while the universe expands for $t>0$ (see figure 2). At $t=0$, the deceleration parameter (which is defined as $q=-\big(1+\frac{\dot{H}}{H^2}\big))$ has a singularity (i.e, $q \rightarrow -\infty$ as $t \rightarrow 0$, (see figure 3)). For $t<0,$ the universe makes a transition from decelerating phase to an accelerating phase while the universe switches over from accelerating phase to matter dominated era at $t>0$. Thus the present quintom model will be in the phantom domain in the neighbourhood of $t=0$ on both sides. On the other hand for the choice $l>Q$, the solution describes the big--bang model of the universe. The big--bang epoch is described by the time instant $t_B=\frac{2}{3z_0} \sqrt{\frac{m(l-Q)}{6}}$ (choosing  $t_0=0$). The universe expands in a power--law manner for $t> t_B$ as in the matter dominated era in standard cosmology. The evolution of the scale factor $'a',$ the Hubble parameter $H$ and the deceleration parameter $'q'$ are also presented graphically in figures (4)--(6) respectively. Note that the other solutions described in equation (59) (or(61)) corresponds to dust era of the standard FLRW universe. Thus in the present quintom cosmological model we have a bouncing solution with a singularity (type IV) [57] in the cosmological parameters (as in the bouncing solutions in the literature). Also, we can describe the phase for $t<0,$ where the universe enters into the present accelerating era from the decelerating epoch, as the observationally supported present state of the universe. \\\\

If future observations predict a lower bound of the equation of state parameter for DE then quintom model may be constrained so that type-IV singularity [57] may be avoided. For future work, the Noether symmetry approach will be used for Minisuperspace Quantum Cosmology, i.e. we shall extend the point transformation to the group--invariant transformations for the Wheeler--Dewitt equation which is equivalent to the existence of conservation laws for the field equations and also indicates the existence of analytical solutions. Further the present symmetry method will be applied to Extended Theories of Gravity. Finally it will be examined how specific Lagrangian multipliers related to symmetries can be used to reduce the dynamics and exact cosmological solutions may be easily obtained.

\section{acknowledgement}
Author SC thanks Inter University Center for Astronomy and Astrophysics (IUCAA), Pune, India for their warm hospitality as a part of the work was done during a visit. Also SC thanks UGC-DRS programme at the Department of Mathematics, Jadavpur University. Author SD thanks Department of Science and Technology (DST), Govt. of India for awarding a Inspire research fellowship. Author ML thanks the Department of Atomic Energy Commission for support in the form of a DAE Raja Ramanna fellowship.


\frenchspacing
\begin{thebibliography}{58}
\bibitem{sj} S. J. Perlmutter et al., {\it Astrophys. J.} {\bf 517}, 565 (1999).
\bibitem{pm} P. M. Garnavich et al., {\it Astrophys. J.} {\bf 493}, L53 (1998).
\bibitem{ag1} A. G. Riess et al., {\it Astron. J} {\bf 116}, 1009 (1998).
\bibitem{ek} E. Komatsu et al.,(WMAP collaboration), {\it Astrophys. J. suppl. ser.} {\bf 180}, 330 (2009).
\bibitem{dn} D. N. Spergel et al.,(WMAP collaboration), {\it Astrophys. J. suppl. ser} {\bf 180}, 330 (2009).
\bibitem{sh} Ho Shirley et al., [{\bf arXiv:} 1201.2137].
\bibitem{wj} W J. Percival et al., [{\bf arXiv:} 0705.3323].
\bibitem{ag2} A. G. Sanchez et al., {\it Mon. Not R. Astron. Soc} {\bf 425}, 415 (2012).
\bibitem{bj} B. Jain and A. Taylor, {\it Phys. Rev. Lett} {\bf 91}, 141302 (2003).
\bibitem{ye} Y. F. Cai, E. N. Saridakis, M. R. Setare and J. Q. Xid, {\it Phys. Rep.} {\bf 493}, 1 (2010).
\bibitem{sc1} S. M. Carroll, AIP Conf.Proc. {\bf 743}, 16 (2004).
\bibitem{ec} E. J. Copland, M. Sami and S. Tsujikawa, {\it Int. J. Mod. Phys. D} {\bf 15}, 1753 (2006).
\bibitem{vs1} V. Sahni and A. Starobinsky, {\it Int. J. Mod. Phys. D} {\bf 15}, 2105 (2006).
\bibitem{sw} S. Weinberg, Rev. Mod. Phys {\bf 61}, 1 (1989).
\bibitem{sc2} S. M. Carroll, W. H. Press and E. L. Turner, {\it Ann. Rev. Astron. Astrophys} {\bf 30}, 499 (1992).
\bibitem{vs2} V. Sahni and A. Starobinsky, {\it Int. J. Mod. Phys. D} {\bf 9}, 377 (2000).
\bibitem{pj} P. J. E.Peebles and B. Ratra, {\it Rev. Mod. Phys.} {\bf 75}, 559 (2003).
\bibitem{tp1} T. Padmanabhan, {\it Phys. Rep.} {\bf 380}, 235 (2003).
\bibitem{tp2} T. Padmanabhan, {\it Curr. Sei.} {\bf 88}, 1057 (2005).
\bibitem{sc3} S. M. Carroll, {\it Liv. Rev. Lett} {\bf 4}, 1 (2001).
\bibitem{la} L. Amendola, S. Tsujikawa, {\it Dark Energy: Theory and Observation. Cambridge univ. press, cambridge} (2010).
\bibitem{rr2} R. R. Caldwell, R. Dave and P. J. Steinhart, {\it Phys. Rev. Lett.} {\bf 80}, 82 (1998).
\bibitem{rr1} R. R. Caldwell, {\it Phys. Lett. B} {\bf 545}, 23 (2002).
\bibitem{jy} J. Yoo and Y. Watanabe, {\it Int. J. Mod. Phys. D} {\bf 21}, 123003 (2012).
\bibitem{pr} P. A. R. Ade et al. [Planck Collaboration], {\it Astron. Astrophys.} {\bf 571}, A22 (2014), [arXiv: 1303.5082].\\
\bibitem{pr1} P. A. R. Ade et al. [{\bf arXiv:} 1502.01589].
\bibitem{jq1} J. Q. Xia, Y. F. Cai, T. T. Qiu, G. B. Zhao and X. Zhang, {\it Int. J. Mod. Phys. D} {\bf 17}, 1229 (2008).
\bibitem{jq2} J. Q. Xia, G. B. Zhao, B. Feng and X. Zhang, {\it J. Cosmol. Astropart. Phys.} {\bf 0609}, 015 (2006).
\bibitem{jq3} G. B. Zhao, J. Q. Xia, B. Feng and X. Zhang, {\it INT. J. Mod. Phys. D} {\bf 16}, 1229 (2007).
\bibitem{jq4} J. Q. Xia, B. Feng and X. Zhang, {\it Mod. Phys. Lett. A} {\bf 20}, 2409 (2005).
\bibitem{bf} B. Feng, M. Li, Y. S. Piao and X. Zhang, {\it Phys. Lett. B} {\bf 634}, 101 (2006).
\bibitem{mr1} M. R. Setare, {\it Phys. Lett. B} {\bf 641}, 130 (2006).
\bibitem{mr2} M. R. Setare, J. Sadeghi and A. R. Amani, {\it Phys. Lett. B} {\bf 660}, 299 (2008).
\bibitem{jq5} J. Q. Xia, Y. F. Cai, T. T. Qiu, G. B. Zhao and X. Zhang, {\it arxiv:} {\bf astro-ph/0703202}.
\bibitem{sr1} S. Capozziello, R. de. Ritis, C. Rubano and P. Scudellaro, {\it Riv. Nuovo Cimento} {\bf 19}, 2 (1996).
\bibitem{ms1} M. Szydlowski et al., {\it Gen. Relt. Grav.} {\bf 38}, 795 (2006).
\bibitem{hg} H. Goldstein, Classical Mechanics 2nd Ed. 1980 {\bf Addison-Wesley, USA}.
\bibitem{sr2} S. Capozziello, A. Stabile and A. Troisi, {\it Class. Quant. Grav.} {\bf 24}, 2153 (2007).
\bibitem{sr3} S. Capozziello, S. Nesseris and L. Perivolaropoulos, {\it J. Cosmol. Astropart. Phys.} {\bf 12}, 009 (2007).
\bibitem{sr4} S. Capozziello and A De Felice, {\it J. Cosmol. Astropart. Phys.} {\bf 08}, 016 (2008).
\bibitem{sr5} S. Capozziello, E. Piedipalumbo, C. Rubano and P. Scudellaro, {\it Phys. Rev. D} {\bf 80}, 104030 (2009).
\bibitem{bv} B. Vakili, {\it Phys. Lett. B.} {\bf 664}, 16 (2008).
\bibitem{yz} Y. Zang, Y. g. Gong and Z. H. Zhu, {\it Phys. Lett. B.} {\bf 688}, 13 (2010).
\bibitem{ms2} M. Szydlowski and M. Heller, {\it Acta. Phys. Pol. B} {\bf 14}, 571 (1983).
\bibitem{ts} M. Tsamparlis and A. Paliathanasis, {\it J. Phys. A} {\bf 44}, 175202 (2011).
\bibitem{pg} P. G. L. Leach, {\it Australian Mathematical Society Lecture Series} {\bf 22}, 12 (2009).
\bibitem{gb} G. Bluman and S. Kumei, {\it Symmetries and Differential Equations} {\bf (Springer-Verlag, N. Y. 1989)};\\
             H. Stephani, {\it Differential Equations: Their Solutions Using Symmetry}
						{\bf (Camb. Univ.Press, Cambridge, England,1989)};\\
						P. J. Olver, {\it Applications Of Lie Groups to Differential Equations} {\bf (Springer, N. Y. 1986)}.
						

\bibitem{aa1} A. V. Aminova, {\it Mat. Sb.} {\bf 186}, 1711 (1995).
\bibitem{aa2} A. V. Aminova and N. A. M. Aminov, {\it Tensor Newser.} {\bf 62}, 65 (2000).
\bibitem{tf} T. Feroze, F. M. Mahomed and A. Qadir, {\it Nonlinear Dynamics} {\bf 45}, 65 (2006).
\bibitem{mt1} M. Tsamparlis and A. Paliathanasis, {\it Gen.Relt.Grav.} {\bf 42}, 2957 (2010).
\bibitem{bs} B. Schutz, {\it Geometrical Methods of Mathematical  Physics} {\bf (Camb. Univ.Press, Cambridge, England,1988)}.
\bibitem{sr6} S. Capozziello, F. Darabi and D. Vernieri, {\it Mod. Phys. Lett. A} {\bf 26}, 65 (2011).

\bibitem{sr7} S. Capozziello and M. De-Laurentis, {\it Int. Geom. Meth. Mod. Phys} {\bf 11}, 1460004 (2014).

\bibitem{emm} E. Noether {\it "Invariante Variationsprobleme," Kgl. Ges. d. Wiss. Nachrichten, Math.-phys. Klasse }, 235 (1918)
\bibitem{mt2} A. Paliathanasis et al., {\it Phys. Rev. D} {\bf 93}, 043528 (2016).\\
A. Paliathanasis et al., {\it Phys. Rev. D} {\bf 91}, 123535 (2015).
\bibitem{ns} S. Nojiri, S. D. Odintsov and S. Tsujikawa, {\it Phys. Rev. D} {\bf 71}, 063004 (2005).

















\end{thebibliography}
\end{document}